# COMPARING OF SWITCHING FREQUENCY ON VECTOR CONTROLLED ASYNCHRONOUS MOTOR

Yılmaz Korkmaz[1], Fatih Korkmaz[2], Ismail Topaloglu[2], Hayati Mamur[2]

[1] Faculty of Technology, Department of Electrical-Electronic Engineering,
Gazi University, Ankara, Turkey
[2] Faculty of Engineering, Department of Electrical-Electronic Engineering,
Çankırı Karatekin University, Çankırı, Turkey

## ABSTRACT

*Nowadays, asynchronous motors have wide range use in many industrial applications. Field oriented control (FOC) and direct torque control (DTC) are commonly used methods in high performance vector control for asynchronous motors. Therefore, it is very important to identify clearly advantages and disadvantages of both systems in the selection of appropriate control methods for many industrial applications. This paper aims to present a new and different perspective regarding the comparison of the switching behaviours on the FOC and the DTC drivers. For this purpose, the experimental studies have been carried out to compare the inverter switching frequencies and torque responses of the asynchronous motor in the FOC and the DTC systems under different working conditions. The dSPACE 1103 controller board was programmed with Matlab/Simulink software. As expected, the experimental studies showed that the FOC controlled motors has a lessened torque ripple. On the other hand, the FOC controlled motor switching frequency has about 65-75% more than the DTC controlled under both loaded and unloaded working conditions.*

## 1. INTRODUCTION

For many years, asynchronous motors have wide range use in industrial applications due to its simple and robust structure and low costs when compared with dc motors. Furthermore, we have better high performance control options now due to development of power electronics in the last few decades. There are two noted control methods in high performance control of asynchronous motors named as: FOC and DTC.

Blaschke proposed the FOC in the 1970s. The FOC was unique and only option that we had on asynchronous motors high performance control until 1980s that the DTC was introduced by Takahashi. Since that day, there have been continual discussions and questions about that: Which one has the best performance on asynchronous motors vector control? [1-3].

The main idea of vector control methods is the control of motor flux and torque components separately like DC motors. The main difference between two methods is that the FOC controls by a rotor or stator field orientation, while the DTC controls by a stator field observation [4]. So, the structural differences in the both control methods are that the FOC uses park transformation, more machine parameters, and current regulators, while the DTC uses Clarke transformation, less machine parameters, and any current regulators. Thus, comparative studies between the two methods show that the DTC is simplicity, a fast dynamic response, and robust against to parameter changes. Despite all these advantages, the DTC also has some handicaps: the most important of them being high torque and current ripples. Evidently, all users have to take into account all these advantages and disadvantages when deciding on which method they will use on designed system [5]. There are some comparative studies regarding both comparison of both the methods, which address the motors speed and torque behaviours. Thanks to this research, it is now known that despite a high torque ripple, the DTC has a fast dynamic response [6-7].

In this paper, a new and different perspective has been presented regarding the comparison of the FOC and the DTC drivers. The voltage source inverter switching frequencies has been compared on both the FOC and the DTC systems. There is no doubt about the importance of switching frequency on power electronic systems because its directly affects switching losses and it means also affects indirectly the efficiency of the drivers. Experimental studies have been carried out to compare switching frequencies and the dSpace 1103 controller board has been programmed with Matlab/Simulink.

## 2. FOC AND DTC METHODS

The FOC is mostly used to control the speed of the motor, not control of the moment, due to its low level sensitivity [8].

In the FOC, the parameters of the motor have to be turned in a d-q reference frame, which consider turns in a synchronous speed (park transformation). Thus, the position of the rotor flux has to be well determined for the success of transformation. In the determining the process of the rotor flux position two basic approaches are used.

Using of flux sensors is the first approach to determine the rotor flux position while to measure the rotor position with an incremental encoder and calculate the angle between the axis of the rotor and the flux, is the second one. Eq. 1 and Eq. 2, describes the calculation of the stator current d-q components.

$$i_{qs}^* = \frac{2}{3}\frac{2}{p}\left(\frac{T_e^*}{\lambda_r^*}\right)\left(\frac{L_r}{L_m}\right) \qquad (1)$$

$$i_{ds}^* = \frac{\lambda_r^*}{L_m} \qquad (2)$$

$i_{qs}^*$ and $i_{ds}^*$ are the stator current d-q components reference values, $\lambda_r^*$ is the rotor flux reference value, and $L_m$ is the mutual inductance, $L_r$ is rotor inductance and p is motor pole pairs value in these equations. The stator three phase currents obtained by using the transformation matrix which is given in Eq. 3.

$$\begin{bmatrix} i_a^* \\ i_b^* \\ i_c^* \end{bmatrix} = \begin{bmatrix} \cos\theta & \sin\theta & 1 \\ \cos\left(\theta - \frac{2\pi}{3}\right) & \sin\left(\theta - \frac{2\pi}{3}\right) & 1 \\ \cos\left(\theta + \frac{2\pi}{3}\right) & \sin\left(\theta + \frac{2\pi}{3}\right) & 1 \end{bmatrix} \begin{bmatrix} i_q^* \\ i_d^* \\ i_0 \end{bmatrix} \quad (3)$$

In Eq. 3., $\theta$ indicates the rotor angular position, and can be obtained from the integration of the sum of the rotor angular speed ($\omega_r$) and the sleep angular speed ($\omega_{sl}$) as given in Eq. 4.

$$\theta = \int (\omega_r + \omega_{sl}) dt \quad (4)$$

and the sleep angular speed is obtained from Eq. 5.

$$\omega_{sl} = \frac{L_m R_r}{L_r \lambda_r^*} i_{qs}^* \quad (5)$$

The FOC control block diagram is created by using Eq.1- Eq.5 and is shown in Fig. 1[9].

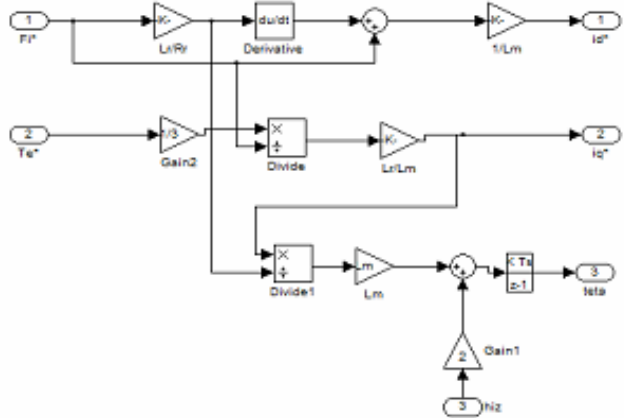

Fig. 1. Simulink block diagram of the FOC controller

The DTC algorithm controls the stator flux and the torque with using measured currents and voltages.

Instantaneous values of the flux and the torque can be obtained by using the transformation of the measured currents and the voltages of the motor. The stator flux is calculated as given in Eq.5-7 in a stationary reference frame [10].

$$\lambda_\alpha = \int (V_\alpha - R_s i_\alpha) dt \quad (6)$$

$$\lambda_\beta = \int (V_\beta - R_s i_\beta) dt \quad (7)$$

$$\lambda = \sqrt{\lambda_\alpha^2 + \lambda_\beta^2} \tag{8}$$

Where, $\lambda_\alpha$-$\lambda_\beta$ stator flux, $i_\alpha$-$i_\beta$ stator current, $V_\alpha$-$V_\alpha$ stator voltage $\alpha-\beta$ components, $R_s$ stator resistance. Motor torque can be calculated as given in Eq.9.

$$T_e = \frac{3}{2} p (\lambda_\alpha i_\beta - \lambda_\beta i_\alpha) \tag{9}$$

Where, *p* is the motor pole pairs. The stator flux vector region is an important parameter for the DTC, and it can be calculated as given in Eq.10.

$$\theta_\lambda = \tan^{-1} \left( \frac{\lambda_\beta}{\lambda_\alpha} \right) \tag{10}$$

The torque and flux errors are obtained by comparing the reference and observed values and than the errors are converted to control signals by hysteresis comparators. The switching table is used to obtain the optimum switching inverter states, and it determines the states by using the hysteresis comparators outputs and the flux region data[11]. The schematic view of the experimental setup is given in Fig. 2.

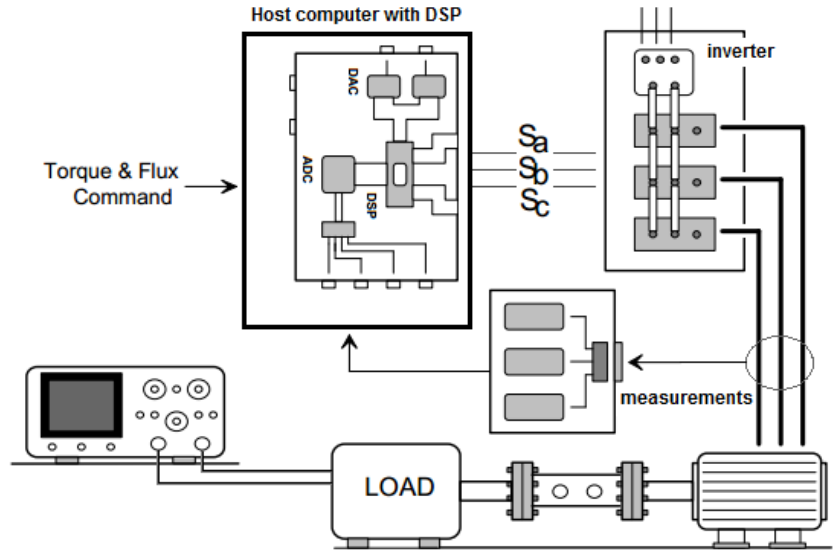

Fig. 2. Schematic view of the experimental setup

## 3. EXPERIMENTAL STUDIES

Experimental tests have been carried out in order to compare the switching frequency between the DTC and the FOC systems, and the dSpace 1103 DSP (Digital Signal Processor) board has been programmed in the Matlab-Simulink Real-Time-Workshop environment for tests.

A new block has been created to compare the switching frequency on both systems. The inverter switching frequency has been obtained by the addition of the arms switching frequencies, and then dividing the total by three. However, especially in the DTC, the switching frequency has a wide range, so the average switching frequency has been calculated with a different sampling time (100 ms) in the frequency measurement block to obtain healthy comparable results.

The results of the experimental tests obtained in this study are for the asynchronous motor of 4 kW and the parameters of the motor, and the experiments are as given below. The motor model is implemented for the DTC and the FOC methods on the Matlab/Simulink. Different load ranges have been applied to the asynchronous motor to compare the switching frequencies. The DTC and the FOC systems have been tested at under no load and loaded (10 N.m.) conditions. The parameters of the three-phase asynchronous motor, in the SI units: P=4 kW, U=380 V, I=8.2 A, cosφ=0.85, 1425 rpm, f=50 Hz, Rs =7.2 Ω.

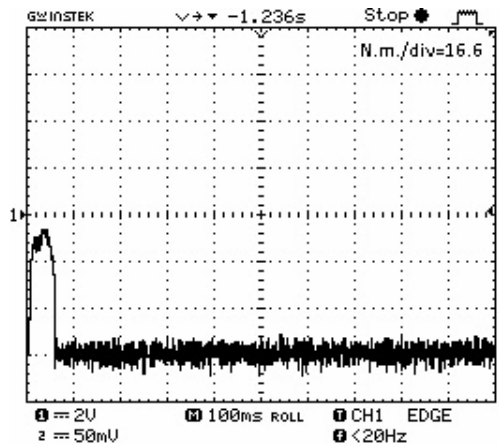

Fig. 3. Torque responses of unloaded DTC controlled motor

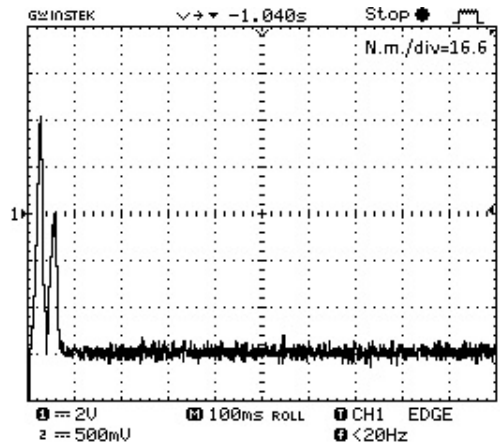

Fig. 4. Torque responses of unloaded FOC controlled motor

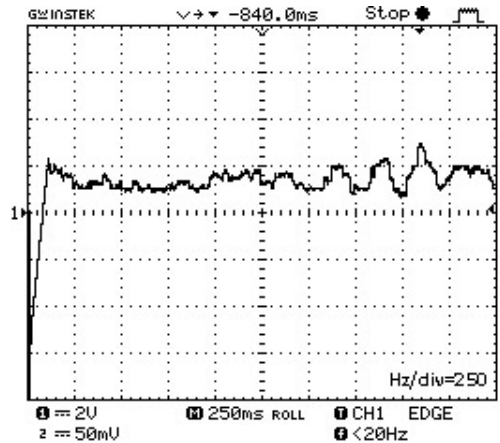

Fig. 5. Switching frequencies of unloaded DTC controlled motor

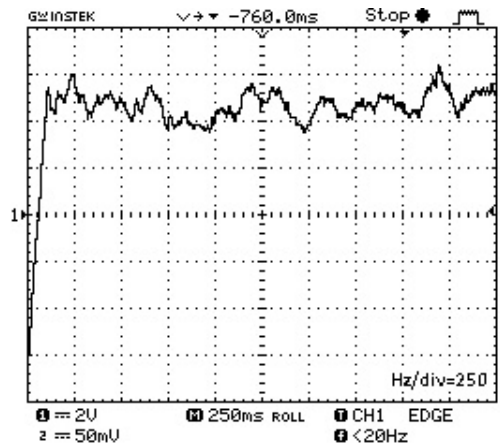

Fig. 6. Switching frequencies of unloaded FOC controlled motor

In the first group of the tests, the motor has been operated as unloaded working conditions. The motor speed has been set up to 1500 rpm for all experimental tests. The moment and switching frequency data, which has been calculated within the Simulink blocks, have been obtained by using the digital-analogue converters of the dSPACE 1103 controller board. The torque response curves of the motor for the both systems have been given in Figure 3 and Figure 4 for unloaded working conditions. As expected, it can be clearly seen that the DTC has much torque ripples when compared FOC.

Figure 5 and Figure 6 show change on the switching frequencies of the motor for both systems. The switching frequency of the DTC controlled motor has nearly 900 Hz, while the FOC controlled motor has 1400 Hz. Therefore, it can be stated that the FOC controlled motor switching frequency has 65% more than the DTC controlled one.

In the second part of tests, the motor has been operated under loaded (10Nm) conditions and torque and the switching frequency curves of these tests have been given in following figures.

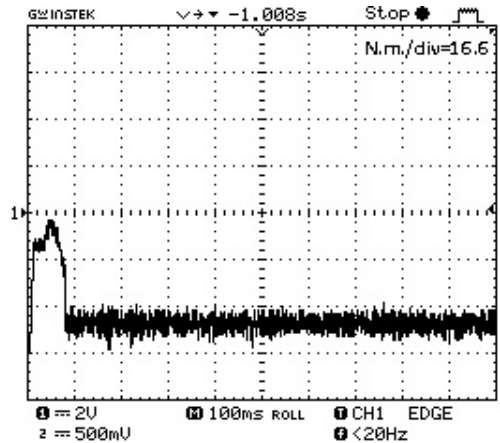

Fig. 7. Torque responses of loaded DTC controlled motor

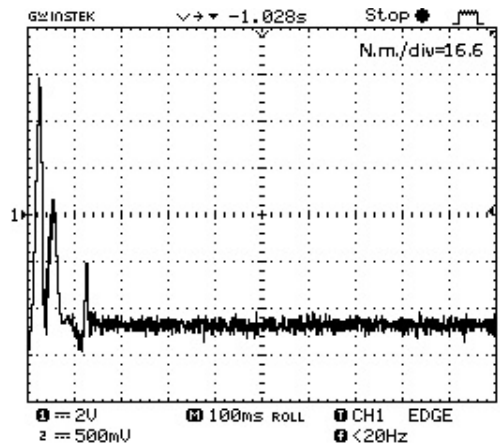

Fig. 8. Torque responses of loaded FOC controlled motor

The torque response curves of the motor for the both systems have been given in Figure 7 and Figure 8 for loaded working conditions. By analyzing the torque responses of the motor, it can still be stated that the DTC has much torque ripple. Figure 9 and Figure 10 show change on the switching frequencies of the motor for both systems. The switching frequency of the DTC controlled motor has approximately 800 Hz, while the FOC controlled motor has 1300 Hz. Consequently, the FOC controlled motor switching frequency nearly %75 more than DTC controlled under loaded conditions.

After all the experimental studies, it can be seen that the switching frequency behaviour of the inverter has close ratio in loaded and unloaded operating conditions, so this shows that the change of the switching frequency on both systems is independent of the motor load situation. In addition, it can be accepted that the change in the switching frequency is also independent of the motor size due to the independency of the load situations.

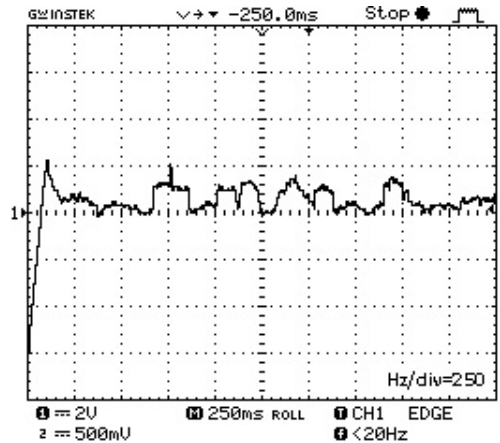

Fig. 9. Switching frequencies of loaded DTC controlled motor

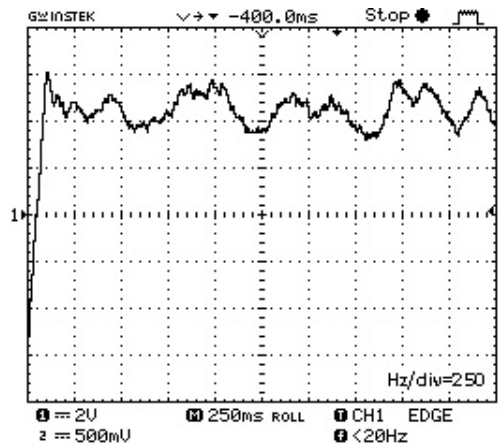

Fig. 10. Switching frequencies of loaded FOC controlled motor

## 4. CONCLUSIONS

Choose of the motor driver can be very important to operate the designed system efficiently when designing an application which includes electric motors. Addition, many times, it can be difficult to decide any optimum control method for obtaining high performance regarding to asynchronous motor drivers. In fact, for the high performance asynchronous motor drivers, we have only two options: FOC and DTC. This paper deals with to give fair comparison between two vector control methods for asynchronous motor drives. For this purpose, experimental tests have been realized to compare of the motor performances on different working conditions. The experimental test shows that the DTC method is preferable if the fast dynamic performance has primary importance whereas the FOC method might be a better option when high torque quality is demanded. In addition, the inverter switching frequencies have been investigated in the FOC and DTC methods to give a new criterion for the selection of optimum control strategy for asynchronous motor high performance control. The experimental studies have carried out to compare the switching frequencies in both methods under different load conditions. As a result, the FOC controlled motor switching frequency was almost 70% more than the DTC controlled one under loaded and unloaded working conditions and choosing of the DTC scheme will also provide high energy efficiency driver if the dynamic performance of the motor has primary importance.